\documentclass[traditabstract]{aa}  

\usepackage{graphicx}	
\usepackage[varg]{txfonts}
\usepackage{longtable}
\usepackage{latexsym}
\usepackage{amssymb}
\usepackage{lscape}
\usepackage[]{natbib}

\begin{document} 

      \title{H$\alpha3$: an H$\alpha$  imaging survey of HI selected galaxies 
      from ALFALFA\thanks{Observations 
      taken at the observatory of San Pedro Martir (Baja California, Mexico), belonging to the
      Mexican Observatorio Astron\'omico Nacional. FITS images are available via
      http://goldmine.mib.infn.it.}
      }
      
      \subtitle{V: The Coma Supercluster survey completion.}

      \author{Giuseppe Gavazzi \inst{1}
      \and Guido Consolandi \inst{1}
      \and Elisa Viscardi \inst{1}
      \and Matteo Fossati \inst{2,3}
      \and Giulia Savorgnan \inst{4}
      \and Michele Fumagalli \inst{5,6}
      \and Leonel Gutierrez \inst{7}
      \and Hector Hernandez Toledo \inst{7}
      \and Alessandro Boselli \inst{8}
      \and Riccardo Giovanelli \inst{9}
      \and Martha P. Haynes \inst{9}
      }

      \authorrunning{G. Gavazzi et al.}
      \titlerunning{H$\alpha3$: H$\alpha$ imaging survey of HI selected galaxies from ALFALFA}

      \institute{Universit\`a degli Studi di Milano-Bicocca, Piazza della Scienza 3, 20126 Milano, Italy\\
      \email {giuseppe.gavazzi@mib.infn.it, guido.consolandi@mib.infn.it, e.viscardi4@campus.unimib.it} 
      \and
      Universit{\"a}ts-Sternwarte M{\"u}nchen, Schenierstrasse 1, D-81679 M{\"u}nchen, Germany. 
      \and
      Max-Planck-Institut f{\"u}r Extraterrestrische Physik, Giessenbachstrasse, D-85748 Garching, Germany\\
      \email {mfossati@mpe.mpg.de}
      \and
      Centre for Astrophysics and Supercomputing, Swinburne University of Technology, Hawthorn, Victoria 3122, Australia\\
      \email gsavorgn@astro.swin.edu.au
      \and
      Institute for Computational Cosmology, Department of Physics, Durham University, South Road, Durham, DH1 3LE, UK
      \and
      Carnegie Observatories, 813 Santa Barbara Street, Pasadena, CA 91101, USA\\
      \email {michele.fumagalli@durham.ac.uk}
      \and
      Instituto de Astronom\'ia, Universidad Nacional Aut\'onoma de M\'exico, 
      Carretera Tijuana-Ensenada, km 103, 22860 Ensenada, B.C., M\'exico.\\
      \email {leonel@astro.unam.mx, hector@astroscu.unam.mx}
      \and
      Aix Marseille Universit\'e, CNRS, LAM (Laboratoire d'Astrophysique de Marseille) UMR 7326, 13388, Marseille, France\\
      \email {alessandro.boselli@oamp.fr}
      \and
      Center for Radiophysics and Space Research, Space Science Building, Ithaca, NY, 14853\\
      \email {haynes@astro.cornell.edu, riccardo@astro.cornell.edu}
      }

      \date{Received; accepted}


   \abstract													     	
    {													     	
     Neutral hydrogen represents the major observable baryonic constituent of galaxies that fuels		       
     the formation of stars through the transformation in molecular hydrogen. 
     The emission of the hydrogen recombination line H$\alpha$
     is the most direct tracer of the process that transforms gas (fuel) into stars.
     We continue to present H$\alpha3$ (acronym for H$\alpha-\alpha\alpha$), an extensive H$\alpha$+[NII] narrow-band imaging campaign of 
     galaxies selected from the HI Arecibo Legacy Fast ALFA Survey (ALFALFA),								       
     using the instrumentation available at the San Pedro Martir observatory (Mexico). 
     In only four years since 2011 we were able to complete in 48 nights the H$\alpha$ imaging observations of 724 galaxies in the
     region of the Coma supercluster $\rm 10^h < R.A. <16^h\, ; \,24^o< Dec. <28^o$ and $3900<cz<9000 \rm ~km~s^{-1}$.
     Of these, 603  are selected from the HI Arecibo Legacy Fast ALFA Survey (ALFALFA) and constitute a 97\% complete sample.
     They provide for the first time a complete census  				
     of the massive star  formation properties of local gas-rich galaxies belonging to
     different environments (cluster vs filaments), morphological type 					       
     (spirals vs dwarf Irr), over a wide range of stellar mass 						       
     ($\sim 10^{8}-10^{11.5}$ M$_\odot$) in the Coma Supercluster.
     The present Paper V provides the H$\alpha$ data and the derived star formation rates for the observed galaxies.
         }

   \maketitle

%

\section{Introduction}

Since the turn of the century broad-band photometry has received a tremendous momentum, in particular from the 
Sloan Digital Sky Survey (SDSS, York et al. 2000).
Owing to this and other similarly large projects that were necessarily carried out by large teams of persons, todays photometric measurements
are available for hundred thousands, if not millions of galaxies in the whole northern sky. \\
Similar extensive work does not exist for narrow-band (eg. H$\alpha$) extragalactic imaging in the nearby universe, and the existing surveys
are still covering small regions of the sky at best.
Besides the growing effort being put in the high-redshift universe, owing to the new generation NIR IFUs attached to 10m class telescopes 
(eg. KMOS, Wisnioski et al. 2014 and SINFONI, F{\"o}rster Schreiber et al.2009 at the VLT; OSIRIS at Keck) 
or narrow-band IR imaging (HiZELS at the UKIRT, Best et al. 2010), 
H$\alpha$ imaging work is available for hardly three thousand galaxies in the local universe.
After the pioneering work of Kennicutt \& Kent, back in 1983, other groups, in collaboration with R. Kennicutt, continued this type of work. Among others
we mention the survey of 468 SINGG galaxies HI selected from HIPASS
by Meurer et al. (2006), the survey of 436 galaxies in the Local Volume (within 11 Mpc) by Kennicutt et al (2008) 
and of 802 objects by Karachentsev \&
Kaisina (2013), and of 465 galaxies 
in Abell clusters by Sakai et al (2012),  approximately 2000 objects in total.\\ 
A similar effort by our group was mostly focused on the Virgo cluster (including parts of the Local Supercluster) and on the Coma supercluster.
It is worth to mention the survey of 482 galaxies in Virgo, Coma and A1367 by Gavazzi et al. (2002a,b), the one of 30 galaxies in the Virgo 
cluster by  Boselli et al. (2002); that of 63 galaxies in Coma+A1367 by Iglesias et al. (2002); the one of 273 galaxies in Virgo+Coma+A1367 by
Gavazzi et al. (2006)  and finally of 235 HI selected galaxies in Virgo by Gavazzi et al. (2012).
Adding the present survey of  724 galaxies in the Coma supercluster, observations by our group add approximately another 1800 H$\alpha$ measurements.

Since the blind HI survey ALFALFA (Giovanelli et al. 2005) was completed (2012) at Arecibo and the catalogue containing 40\% of the targets was
published ($\alpha.40$, Haynes et al. 2011), we undertook the ambitious project to follow-up the HI targets in the spring sky with  H$\alpha$ observations. \\
The project was well suited for the instrumentation available at the San Pedro Martir observatory belonging to the Universidad Nacional
Autonoma de Mexico (UNAM). Both the 1.5m and the 2.1m telescopes are equipped with digital cameras with a field of view of approximately 5 arcmin
and a set of $\sim$ 80 \AA ~wide interferometric filters,
appropriate for covering most targets with pointed observations, reaching the required sensitivity in less than one hour exposure.\\
Paper I of this series (Gavazzi, et. al. 2012) reports the observations obtained for the strip $0<Dec<16$ deg 
(covering 235 galaxies in the Local Supercluster with $cz<3000 \rm ~km~s^{-1}$).
Paper II (Gavazzi, et. al. 2013a) reports the analysis of H$\alpha3$ in the Local Supercluster, 
Paper III (Gavazzi, et. al. 2013b), based on a preliminary analysis of the 
H$\alpha3$ survey in the Great Wall, reports on the evidence for environmental dependent galaxy evolution in the densest
regions of the Coma supercluster. Paper IV (Fossati, et. al. 2013) contains the analysis of the structural parameters 
of H$\alpha3$ galaxies in the Local and Coma Superclusters. \\
The present Paper V contains the data collected for ALFALFA selected galaxies in the strip $24<Dec<28$  deg,
covering  724 objects with  $3900<cz<9000 \rm ~km~s^{-1}$ in the Coma supercluster.\\
The exquisite weather conditions encountered at San Pedro Martir allowed us to complete the Coma survey in just four years (2011-2014).
Since 2011, when H$\alpha3$ began at SPM, the project was allocated with 48 nights. For 257 hours the shutter was kept open.
Counting an average duration of 8 hours per night, this reflects a 67 \% open shutter efficiency.\\ 
This paper is organized as follows. 
The observed Coma supercluster sample is described in Section 2. The observation strategy is the subject of Section 3, while the data reduction procedures are
briefly outlined in Section 4, as they are identical to the ones reported in Paper I. 
The data available for the 724 galaxies are given in 3 tables, whose first pages are given here. 
The full tables, the Atlas and the FITS images of the 724 targeted galaxies are distributed on the WEB via http://goldmine.mib.infn.it/ (Gavazzi et al. 2003, 2014)
under the section $project/papers$. \\
The joined analysis of the Coma supercluster (this paper) and the Local supercluster (Paper I) is the subject of Paper VI of this series
(Gavazzi et al. in prep).\\
Throughout the paper we adopt $H_0=73 \rm ~km~s^{-1}~Mpc^{-1}$.

\section{The Sample}\label{sample}

\subsection{Selection}

\begin{figure*}[!t]
\begin{center}
\includegraphics[scale=0.9, trim= 20 40 10 300]{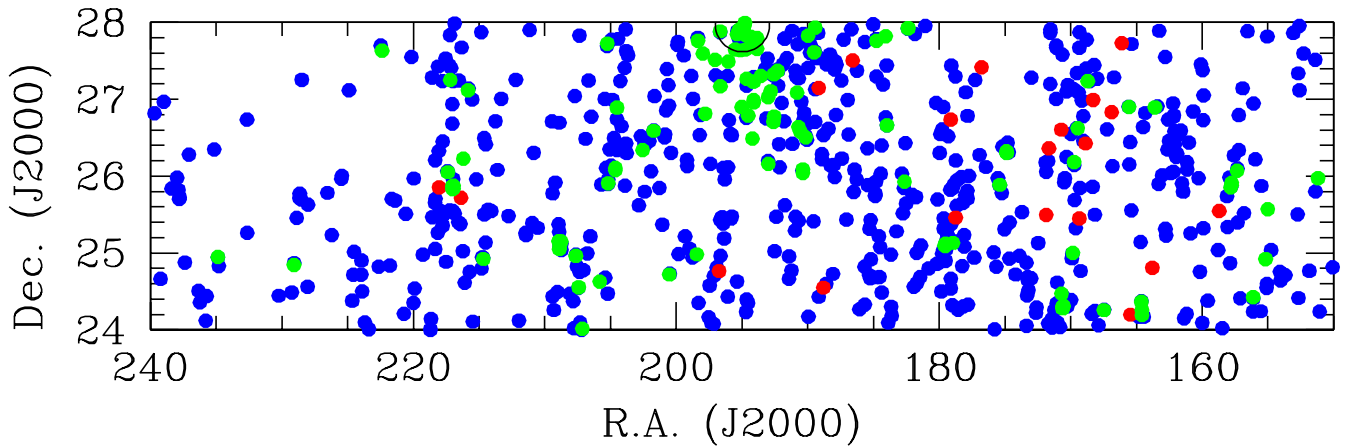}
\end{center}
\caption{Sky distribution (R.A. and Dec. not to scale) of 
623  HI selected galaxies in the ALFALFA strip $24^o<Dec<28^o$; $3900<cz<9000$ $(\rm Km s^{-1})$ (from Haynes et al. 2011),
divided among 603 H$\alpha$ followed-up (blue circles) and 20 not observed (red circles). 
Green circles indicate  121 optically selected galaxies from SDSS without AGC designation. 
The large semi circle gives the position of the Coma cluster.}
\label{fig:Fig1}
\end{figure*}

 Our sample is drawn from the 360 square degree region 
 $\rm 10^h <R.A. <16^h\, ; \,24^o< Dec. <28^o$; $3900<cz<9000\rm ~km~s^{-1}$, 
 covering the Coma Supercluster, including half of the Coma cluster. 
 This region has been fully mapped by ALFALFA (Haynes et al. 2011), which is providing us with a complete sample of HI selected galaxies, 
 with HI masses as low as $10^{9-9.5} ~\rm M_\odot$\footnote{As introduced in Giovanelli et al. (2005)  ALFALFA is
 a noise-limited survey rather than a flux-limited one. At any given integrated HI mass 
 the 21 cm flux per velocity channel is inversely proportional to the width of the HI profile, thus
 to the galaxy inclination.  The completeness and sensitivity of ALFALFA are well understood and discussed
 in detail in  Saintonge (2007), Martin at al. (2010) and Haynes et al. (2011).}. 
 The goal of the H$\alpha3$ survey is to follow up with 
 H$\alpha$ imaging observations the ALFALFA targets with high S/N (typically S/N$~>6.5$)
 and good match between two independent polarizations (code = 1 sources;  (Giovanelli et al. 2005, Haynes et al. 2011).
 We will refer to these targets as the HI or radio targets.
 Fig. \ref{fig:Fig1} illustrates the sky region under study. The panel contains 623 HI selected galaxies in the range 
 $3900<cz<9000$ $\rm (Km s^{-1})$, 
 from the ALFALFA Survey, divided among 603 H$\alpha$ followed-up (blue circles), 20 not observed (red circles),
 plus  121 galaxies (green circles) not ALFALFA-selected (without AGC designation).
 These are for the most part CGCG galaxies previously (1993-2006) observed in the context of the GOLDMine project, independently from their HI selection,
 and some fainter SDSS late type galaxies (LTGs) observed on purpose, especially during the 2013 run,
 for showing  H$\alpha$ emission in their SDSS nuclear spectra, in spite of being gas-poor LTGs (undetected by ALFALFA). This subsample
 serves to test that stripped LT galaxies still retain some gas in their centers feeding some circumnuclear star formation,
 demonstrating that the gas ablation proceeds outside-in, as in the ram-pressure scenario (see Gavazzi et al. 2013b).

\subsection{Completeness}
\label{compl}
 \begin{figure}[!t]
 \includegraphics[width=8cm,height=5cm, trim= 0 0 70 250]{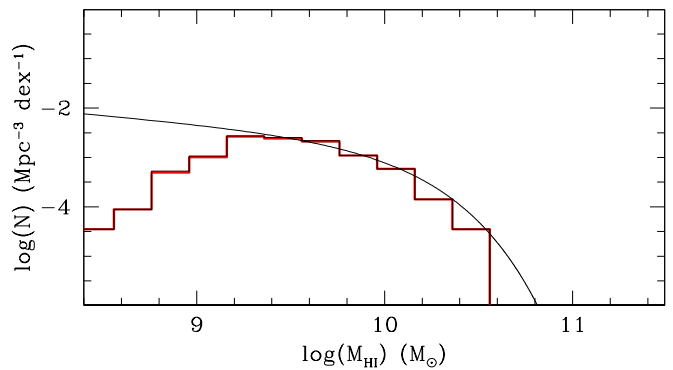}
 \caption{The black solid histogram shows the HI mass function for the 
 623 HI selected galaxies collected from ALFALFA in the range $3900<cz<9000$ $\rm (Km s^{-1})$.
 It is indistinguishable from the red solid histogram, which refers to the H$\alpha$ followed-up galaxies. 
 The black solid line shows the HI mass function of Martin et al. (2010). Notice the lack of bright galaxies above $10^{10.5}$ $M_{\odot}$
 due to the finite sampled volume of the Coma supercluster.}
 \label{fig:compl}
 \end{figure}
 In the region under study there are 683 HI selected galaxies, collected from ALFALFA, but only
 623 of them lie in the range $3900<cz<9000$ $\rm (Km s^{-1})$ (suitable for our narrow-band filter coverage). 603 are followed-up in 
 H$\alpha$ reaching a completeness of 97\%. 
 The remaining 20 ALFALFA sources (red circles in Fig. 1)  have not been observed by H$\alpha3$
 because they lie too close to bright stars that would saturate the detector.  
 
 To further investigate the HI completeness of H$\alpha3$, i.e. the limiting HI mass
 above which H$\alpha3$ is complete,
 we compare in Fig. \ref{fig:compl} the ALFALFA HI mass function
 given by Martin et al. (2010) (black curve),
 representative of the whole Local Universe sampled by ALFALFA,
 which is well represented by a Schechter function with $\alpha$=-1.33, 
 $\Phi_*=4.8\cdot 10^{-3} ~\rm Mpc^{-3}~ dex^{-1}$,
 $M_*=10^{9.96}$
 with the HI Mass distribution in H$\alpha3$ (black solid histogram). 
 The black curve is the ALFALFA HI mass function
 whose $\Phi_*$ has been normalized to the volume sampled by H$\alpha3$.\\
 The agreement between the black line and the black solid histogram is 
 very satisfactory above $\log(M_{\rm HI}/{\rm M_\odot}) \sim 9.2$, 
 that can be assumed as the HI completeness limit of H$\alpha3$.
 The data and the black line depart above $\log(M_{\rm HI}/{\rm M_\odot}) \sim 10.5$ 
 due to cosmic variance as the number of high HI mass galaxies 
 sampled by H$\alpha3$ is limited. This is an effect of HI deficiency 
 and the highest HI mass objects are too rare to be found
 in the small volume sampled by H$\alpha3$. \\
 Fig. \ref{fig:compl} also displays the HI Mass distribution of the 603 H$\alpha$ followed-up galaxies (red histogram).\\

\subsection{Ancillary Data}
\label{ancillary}

 The whole region studied in this paper, where  683 radio selected galaxies are found, has been covered with imaging and 
 spectroscopic observations by the  SDSS (DR10, Ahn et al. 2014), providing for the first time an optically complete selection
 of 2416 galaxies as faint as $r<17.7$ mag.
 It is well known, however that the SDSS pipeline is tailored for providing correct magnitudes of objects of cosmological interest
 ($cz>0.05$), but is often unreliable for extended, low brightness objects in the nearby universe due to
 the "shredding" problem (Blanton et al. 2005).
 Therefore we based our Petrosian magnitude extraction (in the AB system) on the SDSS material using an automated procedure (Consolandi et al. in preparation)
 especially tailored for resolving the shredding problem
 and for accurate masking of the unwanted light coming from contaminating objects (foreground stars and galaxies).
 This procedure performs aperture photometry on the  $i$ and $g$ SDSS images 
 that we downloaded  using the IRSA-Montage software (Katz et al. 2011), centered at the optical coordinates of the target galaxies. 
 The procedure automatically detects and masks (using Sextractor, Bertin \& Arnouts 1996) contaminating  stars and galaxies, evaluates and subtracts the mode value 
 of the sky
 and performs  the integrated photometry up to an aperture centered on each target object corresponding to 
 $2 \times$ the Petrosian radius, consistently with SDSS.
 We checked that the individual measurements obtained using the above procedure are consistent with those 
 measured manually (IRAF) on the SDSS material, even for large galaxies for which the SDSS pipeline is most unreliable.

 The optical properties of the 724 sources observed in H$\alpha3$ are presented in  Table \ref{tab:bas1}.
 Individual entries are as follows:
 \begin{itemize}
 \item{Column  1:} galaxy name using the nomenclature recommended by the IAU;
 \item{Column 2:} AGC designation, from Haynes et al. (2011); 
 \item{Column 3:} CGCG (Zwicky et al. 1968) designation;
 \item{Columns 4 and 5:} optical celestial coordinates (J2000);
 \item{Column 6:} morphological type, classified by the authors by visual inspection of the SDSS color images;
 \item{Column 7:} recessional velocity from the SDSS spectroscopic database or from NED; 
 \item{Columns 8 and 9:} major and minor $25^{th}$ $\rm mag~ arcsec^{-2}$ isophotal diameters in the  ($g$-band) from SDSS; 
 \item{Columns 10 and 11:} $i$ and $g$ Petrosian (AB) magnitudes;  
 \item{Column 12:} adopted distance (Mpc);
 \end{itemize}

\section{Observations}
\label{observ} 

\begin{figure}[!t]
 \centering
 \includegraphics[width=8cm,height=8cm]{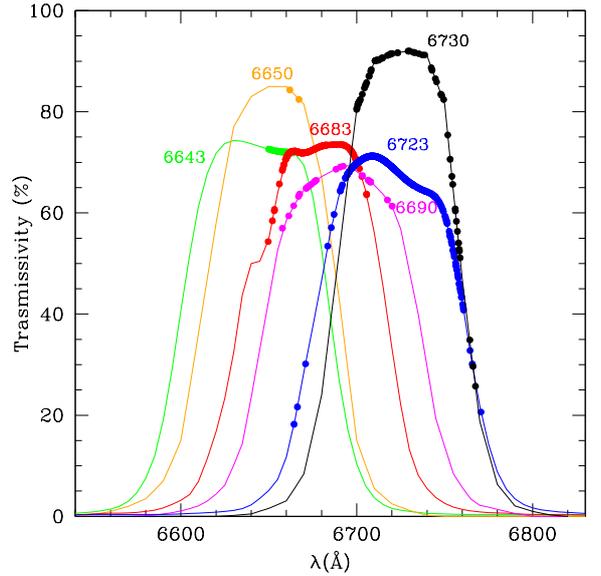}
  \caption{The transmissivity of the ON-band (6643, 6650, 6683, 6690, 6723 and 6730 \AA) filters. 
 Points mark the throughput at the redshift of the target galaxies.} 
 \label{filter}
 \end{figure}

\begin{figure}[!h]
 \centering
 \includegraphics[width=8cm,height=8cm]{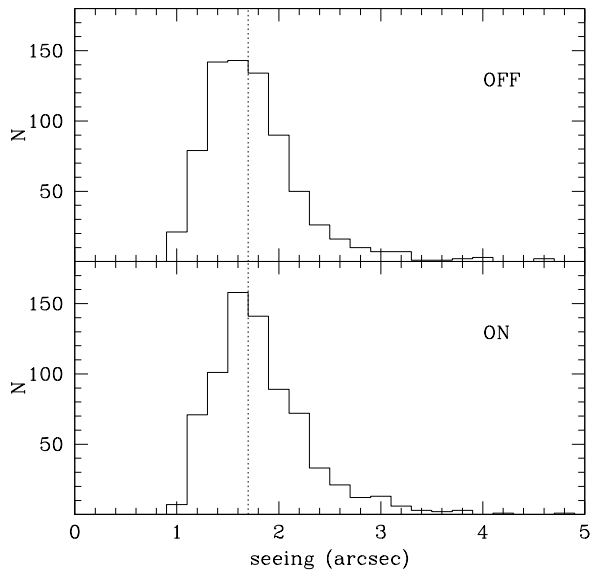}
  \caption{FWHM of stars measured on the final OFF-band images (Top panel) 
  and ON-band images (Bottom panel). Poor telescope guiding performance and dome seeing result in a median seeing
  of 1.7'' (dotted line). 
 }
 \label{fig:see}
 \end{figure}

\begin{table*}[t]
\caption{The log-book of the H$\alpha$ observations in the  Coma region. Prior to 2010 galaxies were optically selected,
while the proper  H$\alpha3$ project (follow-up of ALFALFA) began in 2010.}
\centering
\begin{tabular}{c c c c c c c c}
\hline\hline
  Yr  &	  Date-Obs  &	Telescope &	CCD	  &	 N Pixel	  &	 Rebin &   Pixel scale   &	   N Targets \\
      &             &             &               &		  &	       &   arcsec  &		   \\
\hline  
  1993&  30/4 - 30/4&	SPM2.1    &	TK-1k	  &	 1024x1024&	   1x  &      0.25 &	     2     \\
  1994&  30/3 - 30/3&	SPM2.1    &	TK-1k	  &	 1024x1024&	   1x  &      0.25 &	     1     \\
  1995&  03/4 - 03/4&	SPM2.1    &	TK-1k	  &	 1024x1024&	   1x  &      0.30 &	     4     \\
  1996&  18/4 - 21/4&	SPM2.1    &	TK-1k	  &	 1024x1024&	   1x  &      0.30 &	     9     \\
  1997&  09/3 - 11/3&	SPM2.1    &	TK-1k	  &	 1024x1024&	   1x  &      0.30 &	     12    \\
  1999&  19/4 - 19/4&	SPM2.1    &	TK-1k	  &	 1024x1024&	   1x  &      0.30 &	     2     \\
  2000&  01/3 - 01/3&	INT2.5    &	EEV-2kx4k &	 2048x4096&	   1x  &      0.33 &	     3     \\
  2001&  20/4 - 20/4&	SPM2.1    &	TH-2k	  &	 2048x2048&	   2x  &      0.36 &	     2     \\
  2002&  18/3 - 20/3&	SPM2.1    &	Site3	  &	 1024x1024&	   1x  &      0.31 &	     2     \\
  2004&  12/3 - 18/3&	SPM2.1    &	Site3	  &	 1024x1024&	   1x  &      0.31 &	     4     \\
  2005&  08/4 - 16/4&	SPM2.1    &	Site3	  &	 1024x1024&	   1x  &      0.31 &	     25    \\
  2006&  29/4 - 30/4&	SPM2.1    &	Site3	  &	 1024x1024&	   1x  &      0.31 &	     2     \\
\hline
  2010&  15/4 - 17/4&	SPM1.5    &	e2vm2	  &	 2048x2048&	   2x  &      0.28 &	     8     \\
  2010&  11/5 - 14/5&	SPM1.5    &	e2vm2	  &	 2048x2048&	   2x  &      0.28 &	     14    \\
  2010&  15/4 - 21/4&	SPM2.1    &	TH-2k	  &	 2048x2048&	   2x  &      0.36 &	     27    \\
  2011&  25/3 - 06/4&	SPM1.5    &	Site4	  &	 1024x1024&	   1x  &      0.25 &	     72    \\
  2011&  25/3 - 31/3&	SPM2.1    &	e2vm2	  &	 2048x2048&	   2x  &      0.35 &	     146   \\
  2012&  24/3 - 28/3&	SPM2.1    &	e2vm2	  &	 2048x2048&	   2x  &      0.35 &	     14    \\
  2012&  16/4 - 24/4&	SPM2.1    &	e2vm2	  &	 2048x2048&	   2x  &      0.35 &	     128   \\
  2013&  07/4 - 15/4&	SPM2.1    &	e2vm2	  &	 2048x2048&	   2x  &      0.35 &	     143   \\
  2013&  10/5 - 13/5&	SPM2.1    &	e2vm2	  &	 2048x2048&	   2x  &      0.35 &	     23    \\
  2014&  24/4 - 02/5&	SPM2.1    &	e2vm2	  &	 2048x2048&	   2x  &      0.35 &	     81    \\
\hline
\end{tabular}
\label{table:tab1}
\end{table*}
 Narrow band imaging of the H$\alpha$ line emission (rest frame $\lambda$ = 6562.8 \AA) of 724 galaxies was secured with observations
 taken for the most part at the San Pedro Martir Observatory belonging to the Mexican Observatorio Astron\'omico Nacional (OAN).
 Out of the 724 objects observed, 603 are HI selected from ALFALFA and  121 are optically selected.\\ 
 For each run Table\ref{table:tab1} summarizes the observing dates, the 
 telescope, the characteristics of the CCD detector used, and the number of observed objects. 
  Among the 724 objects  included in Tables \ref {table:tab1}, 68 have been observed in H$\alpha$  prior to 2010 
 (prior to the publication of ALFALFA). These are optically selected galaxies  
 whose data are in common with Gavazzi et al. 
 (1998, 2002a,b, 2006), Boselli \& Gavazzi  (2002), Iglesias-P{\'a}ramo et al. (2002).
 Among them three galaxies were observed in 2000 using the 2.5m Isaac Newton Telescope (INT, La Palma).  \\
 All other target galaxies were observed using the 2.1m and 1.5m telescopes at San Pedro Martir Observatory
 equipped with 1024x1024 pixel detectors from 1993 to 2006 and with a 2048x2048 pixel CCD, used in a 2x rebin mode
 since 2010. \\
 For each galaxy we obtained ON-band exposures using a set of narrow-band interferometric filters, 
 whose bandpass was chosen to include the  wavelength of 
 their redshifted H$\alpha$+[NII] lines, as shown in Fig. \ref{filter}.
  While the median seeing at the San Pedro Martir is $\sim 0''.6$, 
 the final FWHM for point sources in the images is affected by a poor telescope guiding 
 and dome seeing. The final distribution ranges from $\sim 1''$ to $\sim 3''$, 
 with a median seeing $\sigma= 1''.7$ as shown  in Fig. \ref{fig:see}. 
 
 In order to minimize the contamination from cosmic rays 
 we split the ON-band observations in multiple (at least 3) exposures with a total integration time ranging  typically
 from 5 to 30 min, according to the seeing conditions and to the source brightness.\\
 The stellar continuum subtraction was secured by means of shorter (typically 3 to 5 min) 
 observations taken through a broad-band ($\lambda_c$ 6515 \AA, $\Delta\lambda\sim 1200$ \AA) 
 $r$-Gunn filter (OFF-band frames).\\
 We derive the absolute flux calibration using the spectrophotometric stars Feige34 and HZ44 from the catalogue of 
 Massey et al. (1988), observed every $\sim$ 2 hours.
 Most observations were carried out in photometric conditions. However  
 a handful galaxies have been imaged in transparent but not photometric 
 conditions and for these objects we derive only the H$\alpha$ equivalent width 
 ($EW$; insensitive to the absolute flux calibration), but not the H$\alpha$ flux (see N in column 11 of Table \ref{tab:ha1}).
 
 The information relative to the narrow-band observations presented in this paper is listed in Table \ref{tab:obs1}, as follows: 
 \begin{itemize}
 \item{Column  1:} galaxy name using the nomenclature recommended by the IAU;
 \item{Column  2:} AGC designation, from Haynes et al. (2011); 
 \item{Column  3:} CGCG (Zwicky et al. 1968) designation;
 \item{Column  4:} observing date (yy-mm-dd UT);
 \item{Column  5:} central wavelength of the adopted ON-band filter (\AA);
 \item{Columns 6 and 7:} duration and number of individual ON-band exposures;
 \item{Column  8:} average airmass during the ON-band exposures;
 \item{Column  9:} adopted photometric zero point;
 \item{Column  10:} FWHM of point sources (arcsec), as measured on the ON-band frames; 
 \item{Columns 11 and 12:} duration and number of individual OFF-band exposures;
 \item{Column 13:} FWHM of point sources (arcsec) as measured on the OFF-band frames; 
 \item{Column 14:} normalization factor $n$ of the OFF-band frames (see next Section).
 \end{itemize}

\section {Data Reduction}
\label{data} 

\subsection {Image Analysis}  
 We reduce the CCD frames following an identical procedure to the one described in Paper I of this series, 
 based on the STSDAS and GALPHOT IRAF packages. We refer the reader to that paper and give here only a brief 
 summary of the data reduction procedures.
 Also the methods to extract the photometry of the H$\alpha$+[NII] line (flux and equivalent width)
 and to estimate its error budget can be found in Section 4.2 of Paper I. 
 Similarly the procedures to correct the measured flux for Galactic extinction, deblending from [NII] and internal  extinction 
 are identical to the ones given in Section 4.3 of Paper I and are not repeated here.
 
 In summary, each image is bias-subtracted and flat-field corrected using sky exposures obtained during twilight 
 in regions devoid of stars. 
 When three exposures on the same object are available, we adopt a median combination of the realigned images 
 to reject cosmic rays in the final stack. Otherwise we reject cosmic rays by direct inspection of the frames.
 
 We subtract a mean local sky background, 
 computed around the galaxy, using the GALPHOT tasks MARKSKY and SKYFIT.
 Some  frames taken in 2011 are affected by some extra noise structured in a horizontal pattern
 probably introduced by the electronics during the read out process.
 We found that a satisfactory sky background subtraction from these frames
 is obtained using the task BACKGROUND, generally applied to spectroscopic data reduction.
 
 The flat-fielded ON frames were aligned with the OFF frames using field stars. At this stage the seeing was determined
 independently on the two sets of images (see Fig. \ref{fig:see}). After normalization of the OFF-band frames (see Section \ref{photometry}),
 NET images were produced by subtracting the OFF from the ON frames.

\subsection{Integral Photometry}
\label{photometry}

\begin{figure}
 \centering
 \includegraphics[width=8cm,height=8cm]{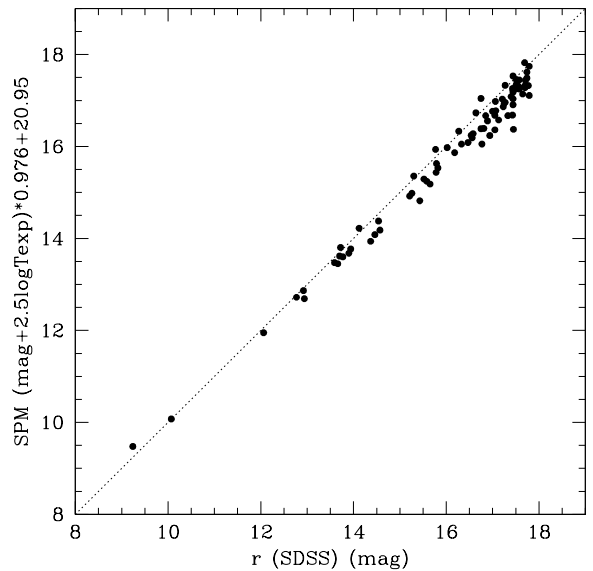}
 \caption{Comparison between the integrated $r$ mag from SDSS (DR10, except for the 10 brightest objects,
 measured as in Section \ref{ancillary})  and the magnitudes measured in the $r$  images taken at SPM in 2014,  using the IRAF task QPHOT.}
 \label{calibR}
 \end{figure}

 Fluxes and $EW$s of the H$\alpha$ line can be recovered from narrow ON-band
 observations by subtracting the stellar continuum contribution estimated from broad-band ($r$) images,
 once these are normalized to account for the ratio of the transmissivity of the two filters
 and the difference in exposure time.
 For each galaxy, we derive the normalization coefficient $n$ by  
 assuming that field stars have no significant H$\alpha$ emission on average
 and therefore they have identical continuum levels in the ON- and OFF-band frames.
 Following  Spector et al. (2012), however we multiply the normalization coefficient found so far by 0.95 
 to account for the fact that field stars are generally redder than the galaxy continua we are trying to estimate. 
   The normalization factors relative to the data taken in 2014 are checked by comparing the $r$-band photometry
  of the target galaxies available from SDSS with the internal magnitudes obtained with aperture photometry  
  performed on our normalized $r$-band frames.
  As shown in Figure \ref{calibR}, a satisfactory agreement exists between the SDSS $r$-band magnitudes and our
  internal magnitudes.

 \subsection{Comparisons with SDSS and the literature}
 
\begin{figure}
 \centering
 \includegraphics[width=8cm,height=8cm]{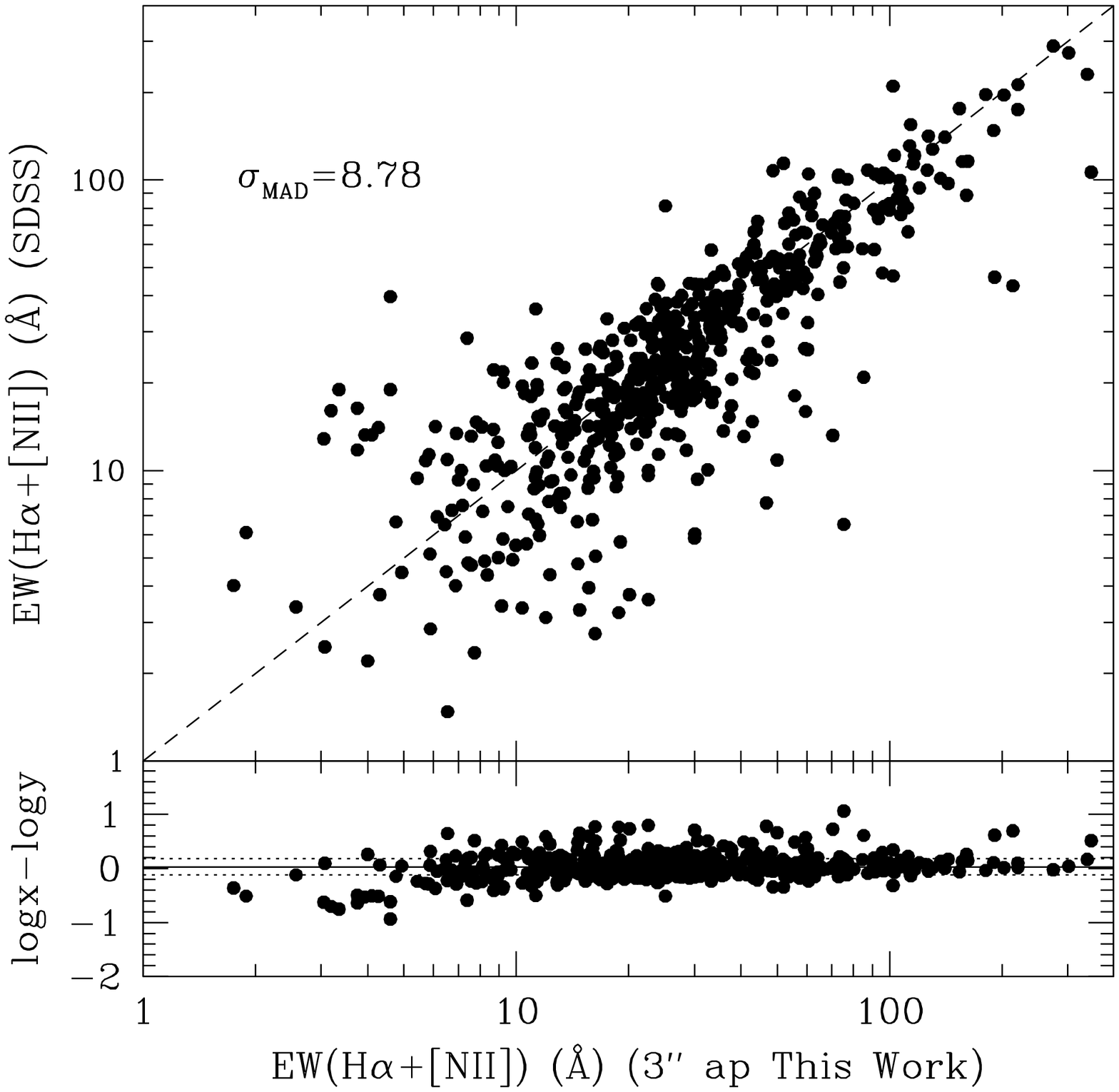}
 \includegraphics[width=8cm,height=8cm]{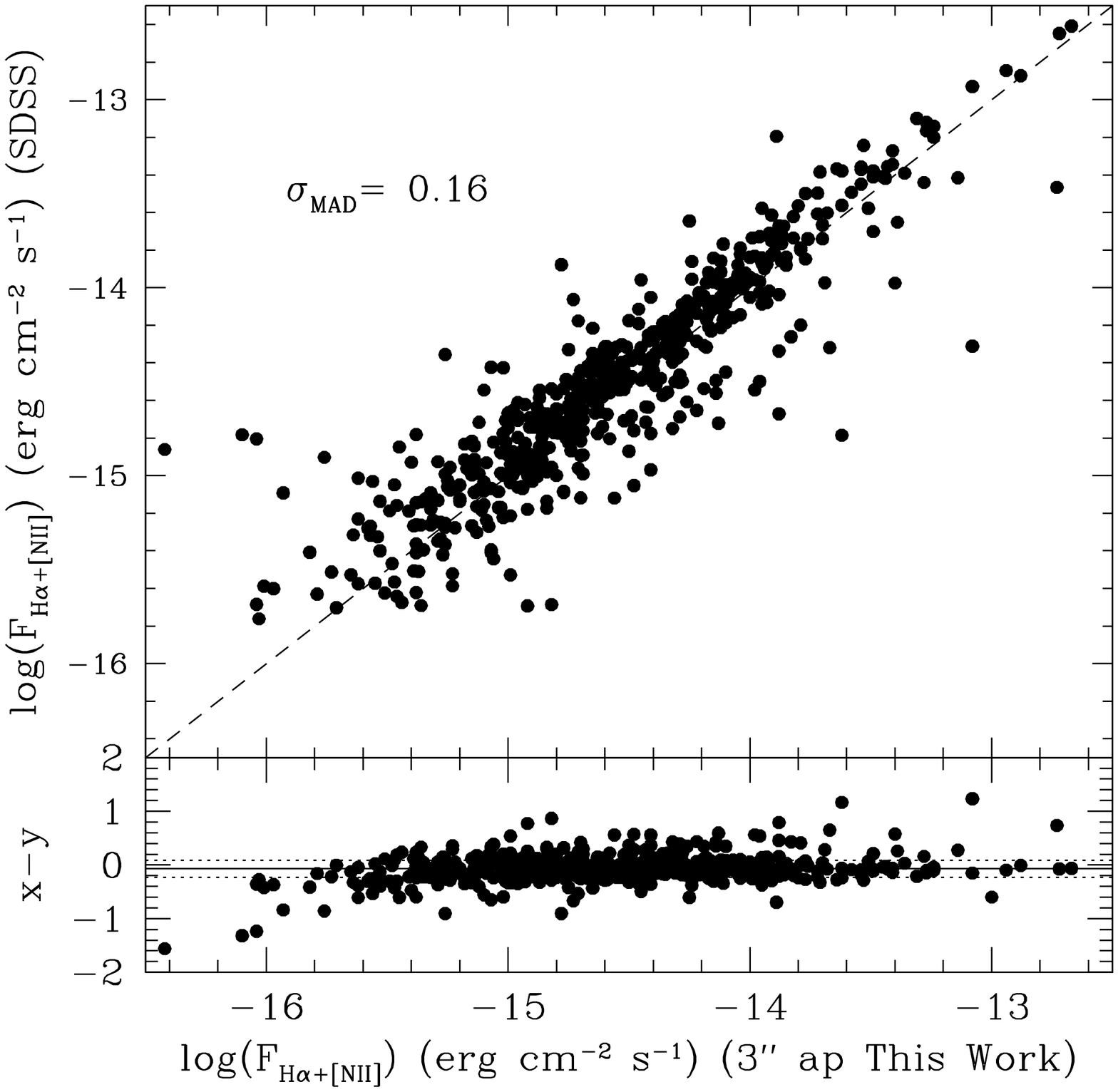}
 \caption{Top panel: comparison between the $EW(\rm H\alpha +[NII])$ measured on the H$\alpha$ images from this work
 in 3 arcsec apertures and the $EW(\rm H\alpha +[NII])$ measured on SDSS spectra taken in 3 arcsec fibres.
 Bottom panel: comparison between the $(\rm H\alpha +[NII])$ flux measured on the H$\alpha$ images from this work
 in 3 arcsec apertures and the $(\rm H\alpha +[NII])$ flux measured on SDSS spectra taken in 3 arcsec fibres.
 The dashed lines give the 1:1 relation.
 }
 \label{compSDSS}
 \end{figure}

 Taking advantage of the SDSS spectral database (DR10), we compare 
 our results, limited to the inner 3 arcsec apertures, with 
 the corresponding values from SDSS nuclear spectra taken in 3 arcsec fibres.\\
 The raw $F(\rm H\alpha +[NII])$ and $EW(\rm H\alpha +[NII])$, (i.e. neglecting corrections
 for [NII] deblending and internal absorption) measured in the inner 3 arcsec aperture centered on the galaxy nuclei 
 are compared with the same quantities from DR10 spectral database 
 in Fig. \ref{compSDSS}, showing overall consistency. 
 The median value of the differences of EW (this work - SDSS) is  1.38 $\pm 8.78 \AA$.
 The median value of the differences of log Flux (this work - SDSS) is -0.07  $\pm 0.16$.\\
 As mentioned in Paper II, the sources of error in the flux measurements are a combination of
 a) uncertainty on the background subtraction (dominant for extended sources),
 b) Poisson statistical uncertainty (dominant for weak sources) and c) systematic uncertainties
 on the OFF-band normalization factor (see Spector et al. 2012).

\begin{figure}
 \centering
 \includegraphics[width=8cm,height=8cm]{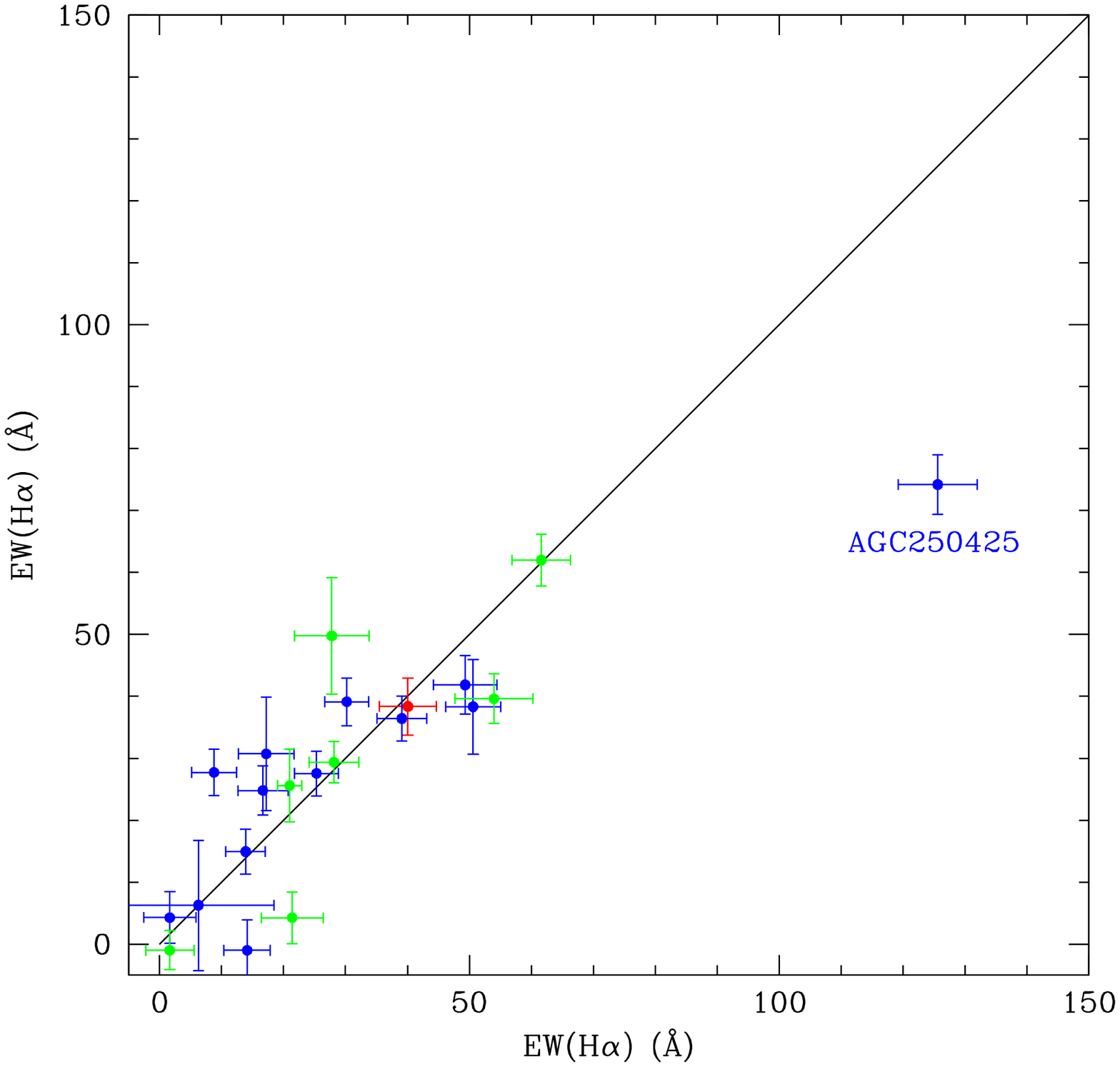}
 \includegraphics[width=8cm,height=8cm]{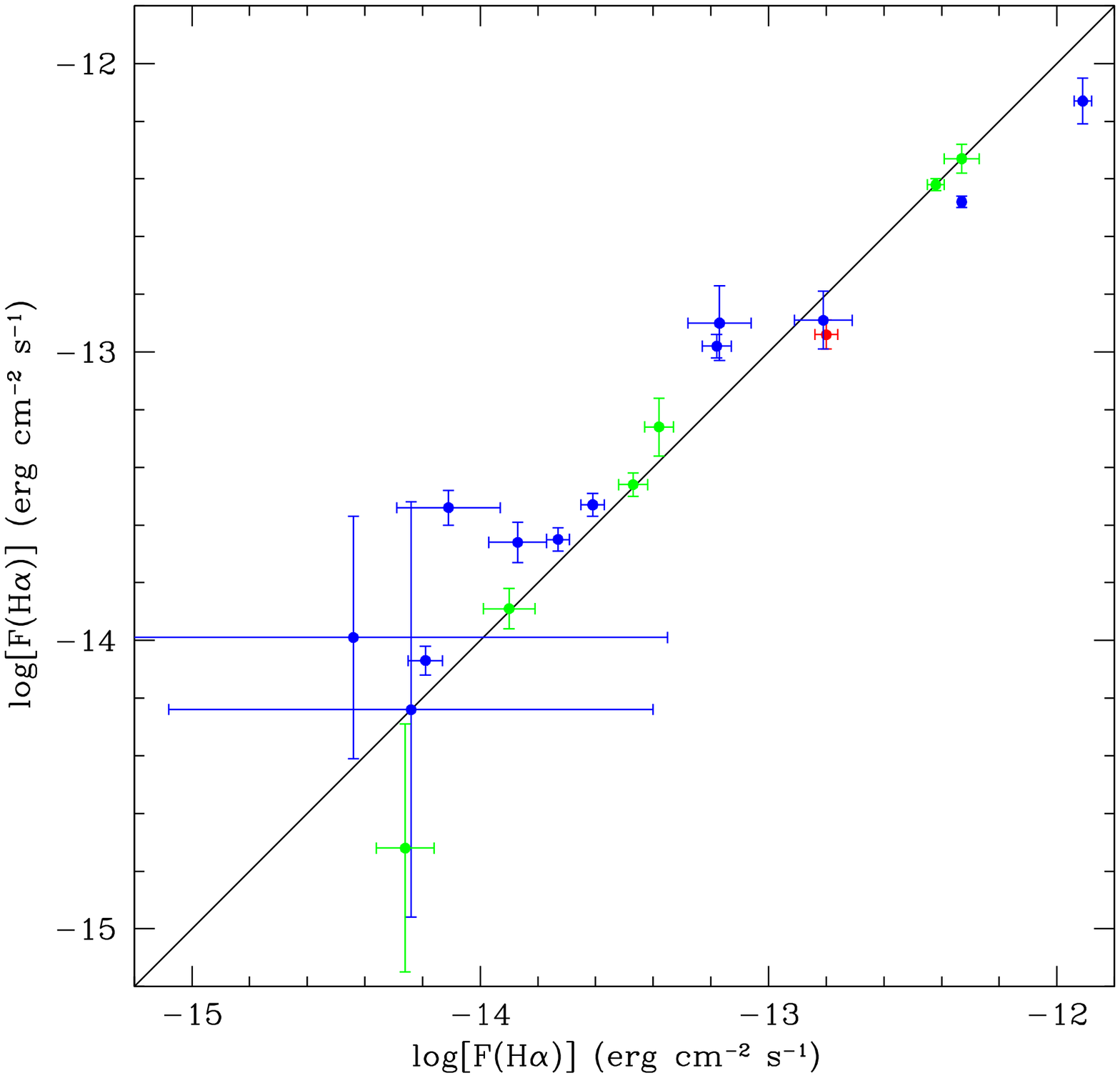}
 \caption{Comparison between the integrated $EW\rm H\alpha$ (top panel) and
 H$\alpha$ flux (bottom panel) repeatedly measured in different observing runs.
 Red symbol refers to measurements taken in April and May 2010 at the 1.5m.
 Blue symbols are 1.5m vs 2.1m measurements.  Green symbols are 2.1m vs 2.1m repeated measurements.
 The dashed lines give the 1:1 relation.
 The object with deviating $EW\rm H\alpha$ was checked on SDSS nuclear spectroscopy
 and found to be consistent with the 2.1m measurement.
 }
 \label{compINT}
 \end{figure}

 Several galaxies were repeatedly observed at SPM in different runs.
 They are given in Fig. \ref{compINT}, showing satisfactory agreement. 
 The only deviating point (AGC 250425) was checked on 3 arcsec aperture and compared 
 with the measurement on the SDSS nuclear spectrum; the correct calibration turns out to correspond 
 to the 2.1m measurement. 

\section{Results}
The results of integral photometry of the H$\alpha$ + [NII] and the derivation of the corrected star formation rate (SFR),
as derived from the present observations, are listed in Table \ref{tab:ha1} as follows:
 
 \begin{itemize}
 \item{Column 1:} galaxy name using the nomenclature recommended by the IAU;
 \item{Column 2:} AGC designation, from Haynes et al. (2011); 
 \item{Column  3:} CGCG (Zwicky et al. 1968) designation;
 \item{Columns 4 and 5:} R.A. and Dec. (J2000);
 \item{Column 6:} equivalent width (EW) of H$\alpha$ + [NII] (\AA);
 \item{Column 7:} 1$\sigma$ uncertainty on the H$\alpha$ + [NII] EW;
 \item{Column 8:} logarithm of H$\alpha$ + [NII]  flux ($\rm erg~cm^{-2}~s^{-1}$);
 \item{Column 9:} logarithm of 1$\sigma$ uncertainty on the H$\alpha$ + [NII] flux;
 \item{Column 10:} logarithm value of SFR in $\rm M_\odot~yr^{-1}$
 corrected for galactic extinction, deblending from [NII] and internal extinction as adopted by Lee et al. (2009);
 \item{Column 11:} sky quality: P = photometric, N = non photometric. An asterisk marks galaxies with uncertain fluxes 
 because the transmissivity of the used ON band filter at their redshift was less than 50\%.
 \end{itemize}

\begin{figure}
 \centering
 \includegraphics[width=8cm,height=8cm]{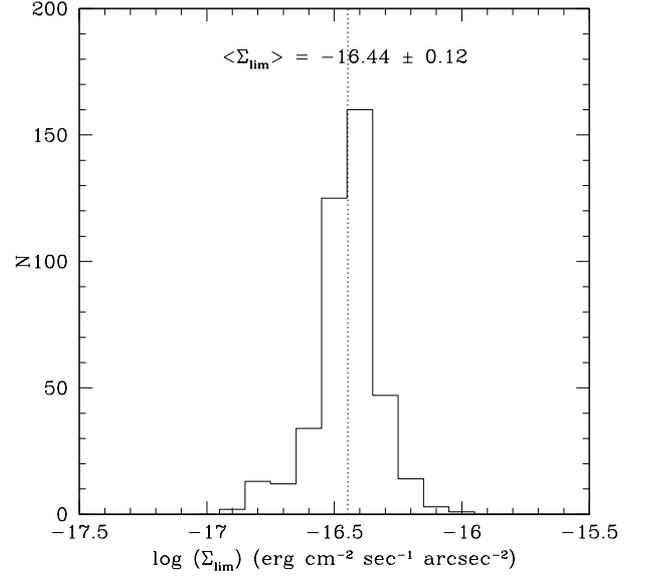}
 \caption{Histogram of the limiting 1$\sigma$ surface brightness in the H$\alpha$ NET images.
 }
 \label{sens}
 \end{figure}

 The limiting sensitivity of the H$\alpha$  observations presented in this paper are given in the histogram of Figure 
 \ref{sens}. The average limiting surface brightness is -16.44 $\pm 0.12$ $(\rm erg ~cm^{-2} ~sec^{-1} ~arcsec^{-2})$.
  
\begin{figure}
 \centering
 \includegraphics[width=12cm,height=12cm]{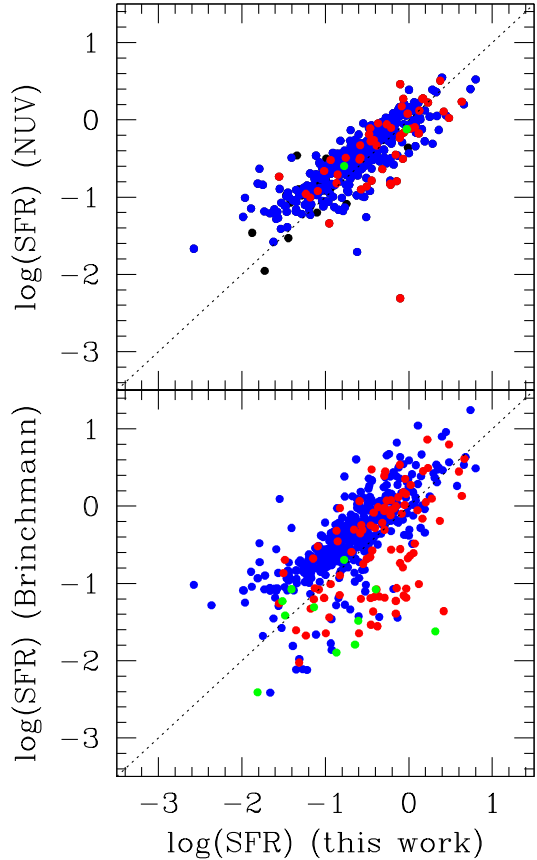}
 \caption{The determination of the star formation rate from this work is compared with the one
 obtained by Brinchmann et al. (2004) (multiplied by 1.5 to account for the change of IMF from Kroupa to Salpeter, bottom panel)  
 and with the one obtained using the NUV magnitudes  of Huang et al (2012) (top panel).
 Blue dots represent galaxies with nuclear HII-like spectra, red dots are AGNs (or LINERs), green dots
 are nuclear passive spectra.
 }
 \label{compSFR}
 \end{figure}

The comparison between the star formation rates determined in this work by converting  H$\alpha$ luminosities
into SFR adopting the Kennicutt (1998) recipe (assuming a Salpeter IMF) with those given by Brinchmann et al. (2004)
and by  Huang et al (2012) is plotted in Figure \ref{compSFR}.
Brinchmann et al. (2004) adopt a Kroupa IMF (Kroupa 2001) and suggest that their SFR should be converted into Salpeter 
by multiplying them by 1.5 (see Figure \ref{compSFR}, bottom panel).
Huang et al (2012) use the Chabrier IMF (Chabrier 2003), therefore we recalculate the SFR from their NUV luminosities 
adopting the Kennicutt (1998) recipe (see Figure \ref{compSFR}, top panel). 
The Figure shows an excellent agreement between our data and Huang et al. (2012). 
Brinchmann et al. (2004) SFRs, based on the SDSS nuclear spectra extrapolated to the whole galaxies using the SDSS integrated colors
appear systematically higher than our values by 0.2 dex. On the opposite most passive and many AGN galaxies 
appear underestimated by Brinchmann et al. (2004). 
Both discrepancies might derive from the extrapolation method adopted by Brinchmann et al. (2004).
Most of the the discrepant AGNs (marked with red symbols) are in fact LINERs, i.e. with  nuclear H$\alpha$ lower or comparable to [NII],
giving a possible clue for their low extrapolated SFR by Brinchmann et al. (2004).

\appendix

\section{The Atlas}\label{atlas}
An Atlas of the 724 observed galaxies, sorted by their celestial coordinates is given in this Appendix.
The OFF-band contours are drawn at 1.5, 2.5, 5 $\times \sigma$ of the sky in the OFF frame and the grey scales represent 
the NET flux intensity between 
1 and 3 $\times \sigma$ of the sky in the NET frame. A bar of one arcmin length is given on all images. 
The images obtained in 2013  have been rotated clockwise by 10.5 or 3.5 degrees to align the Y axes of the CCD with the North direction.

\begin{acknowledgements}
 We thank the Mexican TAC for the generous time allocation to this project.
 We acknowledge useful discussions with Luis Aguillar, Luis Carrasco, Michael Richter.
  We thank Fabrizio Arrigoni Battaia,  Silvia Fabello, Emanuele Farina, Mattia Fumagalli, Lea Giordano and Camilla Pacifici  for their participation in
 some of the observing runs and Luca Cortese, Anna Gallazzi, Stefano Zibetti, Federica Martinelli and Ilaria Arosio for their contribution to the data reduction.\\ 
 The authors would like to acknowledge the work of the entire ALFALFA collaboration team
 in observing, flagging, and extracting the catalog of galaxies used in this work.
 This research has made use of the GOLDMine database (Gavazzi G. et al. 2003, 2014)
 and of the NASA/IPAC Extragalactic Database (NED) which is operated 
 by the Jet Propulsion Laboratory, California Institute of Technology, under contract with the
 National Aeronautics and Space Administration. \\
 Funding for the Sloan Digital Sky Survey (SDSS) and SDSS-II has been provided by the 
 Alfred P. Sloan Foundation, the Participating Institutions, the National Science Foundation, 
 the U.S. Department of Energy, the National Aeronautics and Space Administration, 
 the Japanese Monbukagakusho, and 
 the Max Planck Society, and the Higher Education Funding Council for England. 
 The SDSS Web site is \emph{http://www.sdss.org/}.
 The SDSS is managed by the Astrophysical Research Consortium (ARC) for the Participating Institutions. 
 The Participating Institutions are the American Museum of Natural History, Astrophysical Institute Potsdam, 
 University of Basel, University of Cambridge, Case Western Reserve University, The University of Chicago, 
 Drexel University, Fermilab, the Institute for Advanced Study, the Japan Participation Group, 
 The Johns Hopkins University, the Joint Institute for Nuclear Astrophysics, the Kavli Institute for 
 Particle Astrophysics and Cosmology, the Korean Scientist Group, the Chinese Academy of Sciences (LAMOST), 
 Los Alamos National Laboratory, the Max-Planck-Institute for Astronomy (MPIA), the Max-Planck-Institute 
 for Astrophysics (MPA), New Mexico State University, Ohio State University, University of Pittsburgh, 
 University of Portsmouth, Princeton University, the United States Naval Observatory, and the University 
 of Washington.\\
 R.G. and M.P.H. are supported by US NSF grants AST-1107390 and by a Brinson Foundation grant.
MF acknowledges support by the Science and Technology Facilities Council [grant number ST/L00075X/1].
M Fossati acknowledges the support of the Deutsche Forschungsgemeinschaft via Project ID 387/1-1.
\end{acknowledgements}

\onecolumn


\newpage
\onecolumn
\scriptsize
\tiny
\begin{longtable}{cccccccccccc}
\caption{Basic data of the 724 target galaxies.}\\
\hline\hline
 jName      & AGC      & CGCG		& $R.A.~(J2000)$ & $DEC.~(J2000)$     & Ty  & $cz$	      &  $a$	& $b$	 & $g$   & $i$    & $Dist$ \\
	    &	       &		& hhmmss.ss	 & $^o~'~''$	      &     & ${\rm km~s^{-1}}$ &  arcsec & arcsec &  mag  & mag    & Mpc    \\
 (1)	    & (2)      & (3)		& (4)		 & (5)  	      & (6) & (7)	      & (8)	& (9)   & (10)  & (11)   & (12)   \\
\hline 
\endfirsthead
\caption{continued.}\\
\hline\hline
 jName      & AGC      & CGCG		& $R.A.~(J2000)$ & $DEC.~(J2000)$     & Ty  & $cz$	      &  $a$	& $b$	 & $g$   & $i$    & $Dist$ \\
	    &	       &		& hhmmss.ss	 & $^o~'~''$	      &     & ${\rm km~s^{-1}}$ &  arcsec & arcsec &  mag  & mag    & Mpc    \\
 (1)	    & (2)      & (3)		& (4)		 & (5)  	      & (6) & (7)	      & (8)	& (9)   & (10)  & (11)   & (12)   \\
\hline
\endhead
\hline
\endfoot     
 SDSSJ100016.4+244850	 &  721965  &  -	&  10:00:16.40   &  24:48:50.0    &   Sc    &  6533.7  &     32.81  &	   22.11  &	 16.83  &      16.34  &     89.50  \\	   
 SDSSJ100415.8+241416	 &  721890  &	-	&  10:04:15.88   &  24:14:16.1    &   Sa    &  6334.8  &     49.88  &	   29.08  &	 17.41  &      16.69  &     86.78 \\	   
 SDSSJ100445.2+255832	 &  -	    &	-	&  10:04:45.27   &  25:58:32.4    &   Sa    &  6399.6  &     22.96  &	   12.94  &	 17.27  &      16.34  &     87.67 \\	   
 SDSSJ100531.9+254758	 &  721912  &	-	&  10:05:31.97   &  25:47:58.1    &   BCD   &  5448.9  &     21.57  &	   11.00  &	 16.56  &      15.88  &     74.64 \\	   
 SDSSJ100534.4+273051	 &  5437    &	-	&  10:05:34.48   &  27:30:51.7    &   Pec   &  6382.2  &     69.21  &	   25.58  &	 14.00  &      13.07  &     87.43 \\	   
 SDSSJ100723.0+244558	 &  721938  &	-	&  10:07:23.07   &  24:45:58.5    &   Sm    &  6458.1  &     49.81  &	   9.478  &	 14.27  &      13.27  &     88.47 \\	   
 SDSSJ100909.7+273550	 &  721947  &	-	&  10:09:09.79   &  27:35:50.7    &   Sc    &  6277.2  &     48.86  &	   33.31  &	 17.12  &      16.40  &     85.99 \\	   
 SDSSJ101023.6+270702	 &  721960  &	-	&  10:10:23.64   &  27:07:02.7    &   Sc    &  6417.3  &     56.96  &	   14.97  &	 16.60  &      15.97  &     87.91 \\
\end{longtable}        
\normalsize

\newpage
\scriptsize
\tiny
\begin{longtable}{ccccccccccccccc}
\caption{$H\alpha$ observational specifications of the 724 target galaxies.}\\
\hline\hline
& & & & & \multicolumn{6}{l}{ON}& \multicolumn{4}{l}{OFF}\\ 
\hline
 jName      & AGC      & CGCG	&  Date    &	  Filter       & $T_{\rm exp}$  & $N_{\rm exp}$ & A.M. &  $\log(Zp)$		   & Seeing    & $T_{\rm exp}$ & $N_{\rm exp}$  &   Seeing   & $n$   \\
	    &	       &	&  yymmdd  &	   {\AA}       & sec		&		&      &  $\rm erg~cm^{-2}~s^{-1}$ & arcsec    &  sec	       &		&   arcsec   &       \\
 (1)	    & (2)      & (3)	&  (4)     &	    (5)        & (6)		&  (7)  	& (8)  & (9)			   & (10)      &  (11)         & (12)		&  (13)      & (14)  \\
\hline 
\endfirsthead
\caption{continued.}\\
\hline\hline
& & & & & \multicolumn{6}{l}{ON}& \multicolumn{4}{l}{OFF}\\ 
\hline
 jName      & AGC      & CGCG	&  Date    &	  Filter       & $T_{\rm exp}$  & $N_{\rm exp}$ & A.M. &  $\log(Zp)$		   & Seeing    & $T_{\rm exp}$ & $N_{\rm exp}$  &   Seeing   & $n$   \\
	    &	       &	&  yymmdd  &	   {\AA}       & sec		&		&      &  $\rm erg~cm^{-2}~s^{-1}$ & arcsec    &  sec	       &		&   arcsec   &       \\
 (1)	    & (2)      & (3)	&  (4)     &	    (5)        & (6)		&  (7)  	& (8)  & (9)			   & (10)      &  (11)         & (12)		&  (13)      & (14)  \\
\hline
\endhead
\hline
\endfoot     
  SDSSJ100016.4+244850    &   721965  &   -	   &  2014-04-28  &	   6723        &  600	 &    3  &   1.007   &    -15.36    &	 1.5   &     60    &	3     &   1.5  &   0.80    \\	    
  SDSSJ100415.8+241416    &   721890  &   -	   &  2011-03-27  &	   6730        &  600	 &    3  &   1.098   &    -14.65    &	 1.7   &    300    &	1     &   1.8  &   0.18    \\	    
  SDSSJ100445.2+255832    &   -       &   -	   &  2013-04-08  &	   6723        &  300	 &    3  &   1.006   &    -15.41    &	 1.6   &    60     &	3     &   1.7  &   0.40    \\	    
  SDSSJ100531.9+254758    &   721912  &   -	   &  2012-03-24  &	   6683        &  600	 &    3  &   1.030   &    -15.25    &	 2.1   &    180    &	3     &   2.1  &   0.29    \\	    
  SDSSJ100534.4+273051    &   5437    &   -	   &  2011-03-28  &	   6723        &  600	 &    3  &   1.036   &    -15.35    &	 1.9   &    180    &	1     &   1.6  &   0.15    \\	    
  SDSSJ100723.0+244558    &   721938  &   -	   &  2012-04-17  &	   6723        &  600	 &    3  &   1.033   &       -      &	 1.6   &    60     &	5     &   1.4  &   1.09    \\	    
  SDSSJ100909.7+273550    &   721947  &   -	   &  2011-04-02  &	   6730        &  600	 &    3  &   1.002   &    -14.61    &	 1.1   &    300    &	3     &   1.5  &   0.19    \\	    
  SDSSJ101023.6+270702    &   721960  &   -	   &  2011-04-03  &	   6730        &  600	 &    3  &   1.082   &    -14.47    &	 2.2   &    300    &	1     &   2.1  &   0.17    \\
\end{longtable}
\normalsize

\newpage
\scriptsize
\tiny
\begin{longtable}{ccccccccccc}
\caption{Integrated $H\alpha$ Photometric parameters of the 724 target galaxies.}\\
\hline\hline
 jName      & AGC      & CGCG	&   ${R.A.~(J2000)}$ & ${DEC.~(J2000)}$ &  ${\rm EWH\alpha}$  &  ${\rm \sigma_{EWH\alpha}}$  &  ${\rm \log~F({\rm H\alpha})}$	&  ${\rm \log~\sigma_{F({\rm H\alpha})}}$ &   ${\rm \log~SFR}$        & Quality \\
	    &	       &	&   hhmmss.ss	     & ${^o~'~''}$	&  ${\AA}$	      &  ${\AA}$		     &  ${\rm erg~cm^{-2}~s^{-1}}$	&  ${\rm erg~cm^{-2}~s^{-1}}$		  &   ${\rm M_\odot~y^{-1}}$  & 	\\
 (1)	    & (2)      & (3)	&   (4) 	     & (5)		&  (6)  	      &  (7)			     &  (8)				&  (9) 				  &   (10)		      & (11)	\\
\hline 
\endfirsthead
\caption{continued.}\\
\hline\hline
 jName      & AGC      & CGCG	&   ${R.A.~(J2000)}$ & ${DEC.~(J2000)}$ &  ${\rm EWH\alpha}$  &  ${\rm \sigma_{EWH\alpha}}$  &  ${\rm \log~F({\rm H\alpha})}$	&  ${\rm \log~\sigma_{F({\rm H\alpha})}}$ &   ${\rm \log~SFR}$        & Quality \\
	    &	       &	&   hhmmss.ss	     & ${^o~'~''}$	&  ${\AA}$	      &  ${\AA}$		     &  ${\rm erg~cm^{-2}~s^{-1}}$	&  ${\rm erg~cm^{-2}~s^{-1}}$		  &   ${\rm M_\odot~y^{-1}}$  & 	\\
 (1)	    & (2)      & (3)	&   (4) 	     & (5)		&  (6)  	      &  (7)			     &  (8)				&  (9) 				  &   (10)		      & (11)	\\
\hline 
\endhead
\hline
\endfoot
  SDSSJ100016.4+244850   &  721965   &   -	  &  10:00:16.40   &  24:48:50.0    &	36.80	  &   6.70     &   -13.71   &  0.07    &   -0.802     &     P	\\
  SDSSJ100415.8+241416   &   721890  &   -	  &  10:04:15.88   &  24:14:16.1    &	19.71	  &   5.48     &   -13.30   &  0.11    &   -0.561     &     P	\\
  SDSSJ100445.2+255832   &   -       &   -	  &  10:04:45.27   &  25:58:32.4    &	14.69	  &   3.76     &   -14.13   &  0.11    &   -1.291     &     P	\\
  SDSSJ100531.9+254758   &   721912  &   -	  &  10:05:31.97   &  25:47:58.1    &	45.04	  &   3.76     &   -13.83   &  0.03    &   -1.029     &     P	\\
  SDSSJ100534.4+273051   &   5437    &   -	  &  10:05:34.48   &  27:30:51.7    &	24.73	  &   6.20     &   -13.51   &  0.10    &   -0.695     &     P	\\
  SDSSJ100723.0+244558   &   721938  &   -	  &  10:07:23.07   &  24:45:58.5    &	44.20	  &   7.00     &   -	    &  -       &   -	      &     N	\\
  SDSSJ100909.7+273550   &   721947  &   -	  &  10:09:09.79   &  27:35:50.7    &	11.84	  &  15.10     &   -13.91   &  0.54    &   -1.111     &     P	\\
  SDSSJ101023.6+270702   &   721960  &   -	  &  10:10:23.64   &  27:07:02.7    &	12.26	  &   9.63     &   -13.88   &  0.33    &   -1.041     &     P	\\
\end{longtable} 
 \normalsize

\end{document}